\documentclass{article}

\usepackage{amssymb}
\usepackage{amsfonts}
\usepackage{amsmath}
\usepackage{epsfig}

\usepackage{axodraw}
\usepackage{pstricks}
\usepackage{color}

\usepackage[hmargin=2.50cm,vmargin=2.50cm]{geometry}

\usepackage[utf8]{inputenc}

\begin{document}

\title{Neutrino tri-bi-maximal mixing through sequential dominance}
\author{I. de Medeiros Varzielas \thanks{ivo@cftp.ist.utl.pt} \\
CFTP, Departamento de Física \\
Instituto Superior Técnico \\
Av. Rovisco Pais, 1 \\
1049-001 Lisboa, Portugal
}

\maketitle

\begin{abstract}
We consider an existing framework of models
where the neutrino Dirac and Majorana mass terms are related, 
and clarify how the mass structures interact through the type I seesaw mechanism,
producing exact tri-bi-maximal mixing for the effective neutrinos in the limit of strong Majorana mass hierarchy (sequential dominance).
In order to better illustrate this, we
draw the seesaw diagrams expanded to show the relevant details,
discuss an underlying symmetry that must be preserved by the neutrino mass terms, and
diagonalise the effective neutrino mass matrix corresponding to the sequential dominance limit.
\end{abstract}

\section{Introduction}

The observed neutrino oscillation parameters \cite{PDG, Mariam_data} are well approximated by the
tri-bi-maximal (TBM) mixing structure \cite{Wolfenstein:1978uw, Harrison:2002er, Harrison:2002kp, Harrison:2003aw, Low:2003dz}, in which the leptonic mixing matrix $U_{PMNS}$ takes the form:
\begin{equation}
U_{PMNS} = \left[ 
\begin{array}{ccc}
-\sqrt{\frac{2}{3}} & \sqrt{\frac{1}{3}} & 0 \\ 
\sqrt{\frac{1}{6}} & \sqrt{\frac{1}{3}} & \sqrt{\frac{1}{2}} \\ 
\sqrt{\frac{1}{6}} & \sqrt{\frac{1}{3}} & -\sqrt{\frac{1}{2}}%
\end{array}%
\right]  \label{eq:TBM}
\end{equation}

In \cite{Ivo1}, an $SU(3)_{f}$ family symmetry commuting with an underlying $SO(10)$ supersymmetric (SUSY) grand unified theory (GUT) was used to explain the observed fermion masses and mixings.
In \cite{Ivo3}, within the same SUSY GUT framework of \cite{Ivo1}, $SU(3)_{f}$ was replaced by its discrete subgroup, $\Delta(27)$ \cite{FFK, Muto:1998na, Ma:2006ip, Luhn:2007uq}. The model of \cite{Ivo3} is much simpler and achieves essentially the same results.
Both models feature leptonic mixing close to eq.(\ref{eq:TBM}), obtained through a combination of small mixing angles in the charged lepton sector, and exact TBM mixing in the effective neutrinos.

In this short note we intend to explicitly show how neutrino Dirac and Majorana terms
sharing the same key properties to the terms used in \cite{Ivo1, Ivo3} produce effective neutrino TBM mixing,
when they interact through the type I seesaw mechanism 
\cite{Minkowski:1977sc, Yanagida:1979as, Glashow:1979nm, Mohapatra:1979ia, GellMann:1980vs, Schechter:1980gr}.
One of the necessary ingredients is a strong hierarchy in the Majorana masses,
as the framework relies on sequential dominance \cite{King:1998jw, King:1999cm, King:1999mb, King:2002nf}.

The family symmetry is broken by the vacuum expectation values (VEVs) of the flavon fields. The flavons are Standard Model (SM) singlet fields and are charged under the family symmetry. In the framework of \cite{Ivo1, Ivo3} the flavons are anti-triplets and generically denoted as $\bar{\phi}^{i}$ (the superscript $^{i}$ is the family symmetry index). The fermions are then assigned as triplets of the family symmetry and specifically denoted $\nu_{i}$ and $\nu^{c}_{i}$ (for the neutrinos) or generically denoted as $\psi_{i}$ and $\psi^{c}_{i}$ (respectively, left-handed fermions and the charge conjugate of the right-handed fermions, see \cite{Ivo0}). $H$ represents the usual Higgs fields of the SUSY SM.
Typically terms of the form $\bar{\phi}^{i} \psi_{i} \bar{\phi}^{j} \psi^{c}_{j} H$ will give rise to (Dirac) mass terms through the Froggatt-Nielsen mechanism \cite{fnielsen}, when the flavons and $H$ acquire their VEVs ($\langle \bar{\phi}^{i} \rangle$ and $\langle H \rangle$).

To obtain the required neutrino masses, at least three flavons are used: $\bar{\phi}^{i}_{3}$, $\bar{\phi}^{i}_{23}$ and $\bar{\phi}^{i}_{123}$. The subscript is not an index, but rather a label that is helpful in identifying what VEV each flavon acquires:

\begin{equation}
\left\langle \bar{\phi}_{3}\right\rangle= a
\left(%
\begin{array}{ccc}
0 & 0 & 1%
\end{array}%
\right)
\label{eq:P3 vev}
\end{equation}

\begin{equation}
\left\langle \bar{\phi}_{23}\right\rangle= b
\left(%
\begin{array}{ccc}
0 & 1 & -1%
\end{array}%
\right)
\label{eq:P23 vev}
\end{equation}

\begin{equation}
\left\langle \bar{\phi}_{123}\right\rangle= c
\left(%
\begin{array}{ccc}
1 & 1 & 1%
\end{array}%
\right)
\label{eq:P123 vev}
\end{equation}
The VEVs are in general complex and we omit the phases for simplicity, except the minus in eq.(\ref{eq:P23 vev}) that gives the necessary orthogonality between $\langle \bar{\phi}_{23}\rangle$ and $\langle \bar{\phi}_{123}\rangle$.
These VEVs define a basis in family space, which we denote as the flavon VEV basis. In general, in this basis the charged lepton mass matrix is not diagonal.

In a full model there should be a justification for the non-trivial alignment of these VEVs, and for the absence (or suppression) of some undesirable superpotential terms, but here we simply highlight the role of the VEVs in the framework and outline the desired form of the superpotential. In order to demonstrate how the mixing is obtained it is convenient to focus on the flavon fields that contract with the family index of $\nu_{i}$ and $\nu^{c}_{i}$ in the Yukawa and Majorana superpotential terms, respectively $P_{Y}$ and $P_{M}$.
Additional fields are in general required to make the terms invariant under all the symmetries, but as they contribute solely to the overall magnitude of the term we need not mention them explicitly here, and to keep the discussion focused we will absorb into $\lambda$ coefficients all contributions to a term's magnitude (other than the explicit $a$, $b$, $c$ of eq.(\ref{eq:P3 vev}), eq.(\ref{eq:P23 vev}) and eq.(\ref{eq:P123 vev})). In absorbing magnitude contributions into the $\lambda$, we are also glossing over the Froggatt-Nielsen messengers (for more details on the Froggatt-Nielsen messenger sector see \cite{Ivo0}).

\section{Neutrino Dirac and Majorana mass terms}

We will now illustrate the properties that neutrino masses must possess to belong to the framework of \cite{Ivo1, Ivo3}.

The Yukawa superpotential $P_{Y}$ must have terms that mix $\bar{\phi}^{i}_{23}$ and $\bar{\phi}^{i}_{123}$:

\begin{equation}
P_{Y} = \lambda_{t} P_{t} + \lambda_{@} P_{@} + \lambda_{\odot} P_{\odot}
\label{eq:P_Y}
\end{equation}

\begin{equation}
P_{t} \equiv \bar{\phi}_{3}^{i}\nu_{i}\bar{\phi}_{3}^{j}\nu_{j}^{c} H  \label{eq:P_1}
\end{equation}
\begin{equation}
P_{@} \equiv \bar{\phi}_{23}^{i}\nu_{i}\bar{\phi}_{123}^{j}\nu_{j}^{c} H \label{eq:P_@}
\end{equation}
\begin{equation}
P_{\odot} \equiv \bar{\phi}_{123}^{i}\nu_{i}\bar{\phi}_{23}^{j}\nu_{j}^{c} H \label{eq:P_o}
\end{equation}
As discussed in the introduction, we highlight the structure of the family index contractions by absorbing everything else into the $\lambda$.

In SUSY GUT models (such as \cite{Ivo1, Ivo3}) the term in eq.(\ref{eq:P_1}) also gives rise to the heaviest charged fermion of each family (top quark, tau lepton), so it needs to be greater in magnitude than the other terms:

\begin{equation}
\lambda_{t} \left| a \right| ^{2} > \lambda_{@} \left| b c \right| = \lambda_{\odot} \left| b c \right| \label{eq:Y_hier}
\end{equation}
The equality $\lambda_{@} = \lambda_{\odot}$ (used in \cite{Ivo1, Ivo3}) can be justified by the underlying $SO(10)$ GUT, and plays an important role in getting the tri-maximal mixing.

It is important to note that terms including 
$\bar{\phi}_{23}^{i}\nu_{i}\bar{\phi}_{23}^{j}\nu_{j}^{c}$ and
$\bar{\phi}_{123}^{i}\nu_{i}\bar{\phi}_{123}^{j}\nu_{j}^{c}$ are absent from $P_{Y}$.
In \cite{Ivo1, Ivo3} there are more terms than in eq.(\ref{eq:P_Y}), but they decouple from the neutrino sector.

In contrast with $P_{Y}$, the Majorana superpotential $P_{M}$ needs terms that don't mix 
$\bar{\phi}_{23}^{i}$ and $\bar{\phi}_{123}^{i}$:

\begin{equation}
P_{M} = \lambda_{3} \bar{\phi}_{3}^{i}\nu_{i}^{c} \bar{\phi}_{3}^{j}\nu_{j}^{c} \label{eq:M_3}
\end{equation}
\begin{equation}
+ \lambda_{23} \bar{\phi}_{23}^{i} \nu_{i}^{c}\bar{\phi}_{23}^{j}\nu_{j}^{c} \label{eq:M_23}
\end{equation}
\begin{equation}
+\lambda_{123} \bar{\phi}_{123}^{i} \nu_{i}^{c} \bar{\phi}_{123}^{j} \nu_{j}^{c} \label{eq:M_123}
\end{equation}
Again the $\lambda$ allow us to focus on the family index contractions.

To make use of sequential dominance, the framework relies on the hierarchy of the neutrino Majorana masses:

\begin{equation}
\lambda_{3} \left| a \right| ^{2} \gg \lambda_{23} \left| b \right| ^{2} > \lambda_{123} \left| c \right| ^{2} \label{eq:M_hier}
\end{equation}
The dominance of the term in eq.(\ref{eq:M_3}) is very relevant, as it enables the suppression of the dominant Dirac term, eq.(\ref{eq:P_1}) (through the type I seesaw mechanism). The framework requires the leading Majorana term to feature the contraction of
$\nu_{i}^{c}$ with a field whose VEV aligns in the same direction as 
$\langle \bar{\phi}_{3}^{i} \rangle$ (for eq.(\ref{eq:M_3}) we chose $\bar{\phi}_{3}^{i}$ itself, for simplicity).

It is important to note that terms with family index contractions
$\bar{\phi}_{23}^{i}\nu_{i}^{c}\bar{\phi}_{123}^{j}\nu_{j}^{c}$ are absent from $P_{M}$.

\section{Effective neutrino masses}

\subsection{Expanded seesaw diagrams}

In order to understand how exact TBM mixing is obtained after the type I seesaw, it is useful to analyse expanded seesaw diagrams. We will consider that only right-handed Froggatt-Nielsen messengers contribute (see \cite{Ivo0} for details), so we draw diagrams with $\langle H \rangle$ next to $\nu$ and neglect all diagrams with $\langle H \rangle$ in intermediate positions.

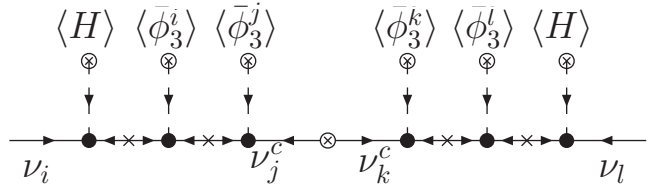
\begin{figure}[!h]
\begin{center}
\fcolorbox{white}{white}{
\begin{picture}(255,76) (45,-299)
    \SetWidth{0.5}
    \SetColor{Black}
    \ArrowLine(45,-269)(75,-269)
    \Vertex(75,-269){2.83}
    \Vertex(105,-269){2.83}
    \Vertex(135,-269){2.83}
    \Line(88,-271)(92,-267)\Line(88,-267)(92,-271)
    \Line(118,-271)(122,-267)\Line(118,-267)(122,-271)
    \ArrowLine(165,-269)(135,-269)
    \ArrowLine(165,-269)(195,-269)
    \COval(165,-269)(2.83,2.83)(45.0){Black}{White}\Line(163.59,-270.41)(166.41,-267.59)\Line(163.59,-267.59)(166.41,-270.41)
    \COval(75,-239)(2.83,2.83)(45.0){Black}{White}\Line(73.59,-240.41)(76.41,-237.59)\Line(73.59,-237.59)(76.41,-240.41)
    \COval(105,-239)(2.83,2.83)(45.0){Black}{White}\Line(103.59,-240.41)(106.41,-237.59)\Line(103.59,-237.59)(106.41,-240.41)
    \COval(135,-239)(2.83,2.83)(45.0){Black}{White}\Line(133.59,-240.41)(136.41,-237.59)\Line(133.59,-237.59)(136.41,-240.41)
    \COval(195,-239)(2.83,2.83)(45.0){Black}{White}\Line(193.59,-240.41)(196.41,-237.59)\Line(193.59,-237.59)(196.41,-240.41)
    \COval(225,-239)(2.83,2.83)(45.0){Black}{White}\Line(223.59,-240.41)(226.41,-237.59)\Line(223.59,-237.59)(226.41,-240.41)
    \COval(255,-239)(2.83,2.83)(45.0){Black}{White}\Line(253.59,-240.41)(256.41,-237.59)\Line(253.59,-237.59)(256.41,-240.41)
    \Vertex(195,-269){2.83}
    \Vertex(225,-269){2.83}
    \Vertex(255,-269){2.83}
    \ArrowLine(285,-269)(255,-269)
    \Line(208,-271)(212,-267)\Line(208,-267)(212,-271)
    \Line(238,-271)(242,-267)\Line(238,-267)(242,-271)
    \ArrowLine(240,-269)(255,-269)
    \ArrowLine(240,-269)(225,-269)
    \ArrowLine(210,-269)(225,-269)
    \ArrowLine(210,-269)(195,-269)
    \ArrowLine(120,-269)(135,-269)
    \ArrowLine(120,-269)(105,-269)
    \ArrowLine(90,-269)(75,-269)
    \ArrowLine(90,-269)(105,-269)
    \DashArrowLine(75,-239)(75,-269){5}
    \DashArrowLine(105,-239)(105,-269){5}
    \DashArrowLine(135,-239)(135,-269){5}
    \DashArrowLine(195,-239)(195,-269){5}
    \DashArrowLine(225,-239)(225,-269){5}
    \DashArrowLine(255,-239)(255,-269){5}
    \Text(50,-286)[lb]{\Large{\Black{$\nu_{i}$}}}
    \Text(268,-286)[lb]{\Large{\Black{$\nu_{l}$}}}
    \Text(136,-286)[lb]{\Large{\Black{$\nu^{c}_{j}$}}}
    \Text(177,-286)[lb]{\Large{\Black{$\nu^{c}_{k}$}}}
    \Text(62,-235)[lb]{\Large{\Black{$\langle H \rangle$}}}
    \Text(92,-235)[lb]{\Large{\Black{$\langle \bar{\phi}_{3}^{i} \rangle$}}}
    \Text(122,-235)[lb]{\Large{\Black{$\langle \bar{\phi}_{3}^{j} \rangle$}}}
    \Text(182,-235)[lb]{\Large{\Black{$\langle \bar{\phi}_{3}^{k} \rangle$}}}
    \Text(212,-235)[lb]{\Large{\Black{$\langle \bar{\phi}_{3}^{l}\rangle$}}}
    \Text(242,-235)[lb]{\Large{\Black{$\langle H \rangle$}}}
  \end{picture}
}
\caption{$P_{t}$ decouples from the effective neutrinos.}
\label{fig:P_1}
\end{center}
\end{figure}

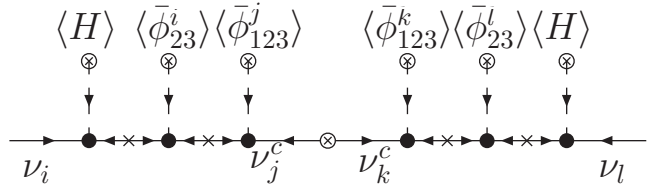
\begin{figure}[!h]
\begin{center}
\fcolorbox{white}{white}{
\begin{picture}(255,76) (45,-299)
    \SetWidth{0.5}
    \SetColor{Black}
    \ArrowLine(45,-269)(75,-269)
    \Vertex(75,-269){2.83}
    \Vertex(105,-269){2.83}
    \Vertex(135,-269){2.83}
    \Line(88,-271)(92,-267)\Line(88,-267)(92,-271)
    \Line(118,-271)(122,-267)\Line(118,-267)(122,-271)
    \ArrowLine(165,-269)(135,-269)
    \ArrowLine(165,-269)(195,-269)
    \COval(165,-269)(2.83,2.83)(45.0){Black}{White}\Line(163.59,-270.41)(166.41,-267.59)\Line(163.59,-267.59)(166.41,-270.41)
    \COval(75,-239)(2.83,2.83)(45.0){Black}{White}\Line(73.59,-240.41)(76.41,-237.59)\Line(73.59,-237.59)(76.41,-240.41)
    \COval(105,-239)(2.83,2.83)(45.0){Black}{White}\Line(103.59,-240.41)(106.41,-237.59)\Line(103.59,-237.59)(106.41,-240.41)
    \COval(135,-239)(2.83,2.83)(45.0){Black}{White}\Line(133.59,-240.41)(136.41,-237.59)\Line(133.59,-237.59)(136.41,-240.41)
    \COval(195,-239)(2.83,2.83)(45.0){Black}{White}\Line(193.59,-240.41)(196.41,-237.59)\Line(193.59,-237.59)(196.41,-240.41)
    \COval(225,-239)(2.83,2.83)(45.0){Black}{White}\Line(223.59,-240.41)(226.41,-237.59)\Line(223.59,-237.59)(226.41,-240.41)
    \COval(255,-239)(2.83,2.83)(45.0){Black}{White}\Line(253.59,-240.41)(256.41,-237.59)\Line(253.59,-237.59)(256.41,-240.41)
    \Vertex(195,-269){2.83}
    \Vertex(225,-269){2.83}
    \Vertex(255,-269){2.83}
    \ArrowLine(285,-269)(255,-269)
    \Line(208,-271)(212,-267)\Line(208,-267)(212,-271)
    \Line(238,-271)(242,-267)\Line(238,-267)(242,-271)
    \ArrowLine(240,-269)(255,-269)
    \ArrowLine(240,-269)(225,-269)
    \ArrowLine(210,-269)(225,-269)
    \ArrowLine(210,-269)(195,-269)
    \ArrowLine(120,-269)(135,-269)
    \ArrowLine(120,-269)(105,-269)
    \ArrowLine(90,-269)(75,-269)
    \ArrowLine(90,-269)(105,-269)
    \DashArrowLine(75,-239)(75,-269){5}
    \DashArrowLine(105,-239)(105,-269){5}
    \DashArrowLine(135,-239)(135,-269){5}
    \DashArrowLine(195,-239)(195,-269){5}
    \DashArrowLine(225,-239)(225,-269){5}
    \DashArrowLine(255,-239)(255,-269){5}
    \Text(50,-286)[lb]{\Large{\Black{$\nu_{i}$}}}
    \Text(268,-286)[lb]{\Large{\Black{$\nu_{l}$}}}
    \Text(136,-286)[lb]{\Large{\Black{$\nu^{c}_{j}$}}}
    \Text(177,-286)[lb]{\Large{\Black{$\nu^{c}_{k}$}}}
    \Text(62,-235)[lb]{\Large{\Black{$\langle H \rangle$}}}
    \Text(92,-235)[lb]{\Large{\Black{$\langle \bar{\phi}_{23}^{i} \rangle$}}}
    \Text(122,-235)[lb]{\Large{\Black{$\langle \bar{\phi}_{123}^{j} \rangle$}}}
    \Text(178,-235)[lb]{\Large{\Black{$\langle \bar{\phi}_{123}^{k} \rangle$}}}
    \Text(212,-235)[lb]{\Large{\Black{$\langle \bar{\phi}_{23}^{l}\rangle$}}}
    \Text(242,-235)[lb]{\Large{\Black{$\langle H \rangle$}}}
  \end{picture}
}
\caption{Diagram giving mass to effective neutrino eigenstate $\nu_{@}$.}
\label{fig:P_@}
\end{center}
\end{figure}

\begin{figure}[!h]
\begin{center}
\fcolorbox{white}{white}{
\begin{picture}(255,76) (45,-299)
    \SetWidth{0.5}
    \SetColor{Black}
    \ArrowLine(45,-269)(75,-269)
    \Vertex(75,-269){2.83}
    \Vertex(105,-269){2.83}
    \Vertex(135,-269){2.83}
    \Line(88,-271)(92,-267)\Line(88,-267)(92,-271)
    \Line(118,-271)(122,-267)\Line(118,-267)(122,-271)
    \ArrowLine(165,-269)(135,-269)
    \ArrowLine(165,-269)(195,-269)
    \COval(165,-269)(2.83,2.83)(45.0){Black}{White}\Line(163.59,-270.41)(166.41,-267.59)\Line(163.59,-267.59)(166.41,-270.41)
    \COval(75,-239)(2.83,2.83)(45.0){Black}{White}\Line(73.59,-240.41)(76.41,-237.59)\Line(73.59,-237.59)(76.41,-240.41)
    \COval(105,-239)(2.83,2.83)(45.0){Black}{White}\Line(103.59,-240.41)(106.41,-237.59)\Line(103.59,-237.59)(106.41,-240.41)
    \COval(135,-239)(2.83,2.83)(45.0){Black}{White}\Line(133.59,-240.41)(136.41,-237.59)\Line(133.59,-237.59)(136.41,-240.41)
    \COval(195,-239)(2.83,2.83)(45.0){Black}{White}\Line(193.59,-240.41)(196.41,-237.59)\Line(193.59,-237.59)(196.41,-240.41)
    \COval(225,-239)(2.83,2.83)(45.0){Black}{White}\Line(223.59,-240.41)(226.41,-237.59)\Line(223.59,-237.59)(226.41,-240.41)
    \COval(255,-239)(2.83,2.83)(45.0){Black}{White}\Line(253.59,-240.41)(256.41,-237.59)\Line(253.59,-237.59)(256.41,-240.41)
    \Vertex(195,-269){2.83}
    \Vertex(225,-269){2.83}
    \Vertex(255,-269){2.83}
    \ArrowLine(285,-269)(255,-269)
    \Line(208,-271)(212,-267)\Line(208,-267)(212,-271)
    \Line(238,-271)(242,-267)\Line(238,-267)(242,-271)
    \ArrowLine(240,-269)(255,-269)
    \ArrowLine(240,-269)(225,-269)
    \ArrowLine(210,-269)(225,-269)
    \ArrowLine(210,-269)(195,-269)
    \ArrowLine(120,-269)(135,-269)
    \ArrowLine(120,-269)(105,-269)
    \ArrowLine(90,-269)(75,-269)
    \ArrowLine(90,-269)(105,-269)
    \DashArrowLine(75,-239)(75,-269){5}
    \DashArrowLine(105,-239)(105,-269){5}
    \DashArrowLine(135,-239)(135,-269){5}
    \DashArrowLine(195,-239)(195,-269){5}
    \DashArrowLine(225,-239)(225,-269){5}
    \DashArrowLine(255,-239)(255,-269){5}
    \Text(50,-286)[lb]{\Large{\Black{$\nu_{i}$}}}
    \Text(268,-286)[lb]{\Large{\Black{$\nu_{l}$}}}
    \Text(136,-286)[lb]{\Large{\Black{$\nu^{c}_{j}$}}}
    \Text(177,-286)[lb]{\Large{\Black{$\nu^{c}_{k}$}}}
    \Text(62,-235)[lb]{\Large{\Black{$\langle H \rangle$}}}
    \Text(88,-235)[lb]{\Large{\Black{$\langle \bar{\phi}_{123}^{i} \rangle$}}}
    \Text(122,-235)[lb]{\Large{\Black{$\langle \bar{\phi}_{23}^{j} \rangle$}}}
    \Text(178,-235)[lb]{\Large{\Black{$\langle \bar{\phi}_{23}^{k} \rangle$}}}
    \Text(208,-235)[lb]{\Large{\Black{$\langle \bar{\phi}_{123}^{l}\rangle$}}}
    \Text(242,-235)[lb]{\Large{\Black{$\langle H \rangle$}}}
  \end{picture}
}
\caption{Diagram giving mass to effective neutrino eigenstate $\nu_{\odot}$.}
\label{fig:P_o}
\end{center}
\end{figure}
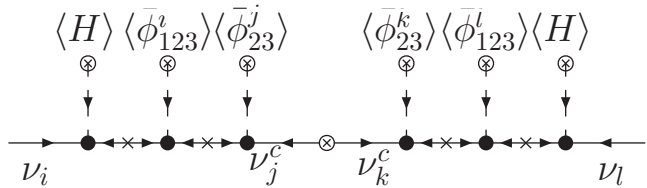

In order to obtain the effective neutrino masses, we start with one of the $P_{Y}$ terms, which converts a $\nu$ into a $\nu^{c}$, then the $\nu^{c}$ have Majorana masses from $P_{M}$, and $\nu$ appears in the external line again through a term from $P_{Y}$.
Starting with eq.(\ref{eq:P_1}), eq.(\ref{eq:M_3}) is the only Majorana term involving the necessary $\bar{\phi}_{3}^{i} \nu^{c}_{i}$ contraction\footnote{The requirement is that $\nu^{c}_{i}$ must contract with a field whose VEV is aligned along $\langle \bar{\phi}_{3}^{i} \rangle$, not necessarily $\bar{\phi}_{3}^{i}$ itself (as in \cite{Ivo1, Ivo3} where a different field contracts with $\nu^{c}_{i}$).},
and we conclude the diagram with the only Yukawa term involving $\bar{\phi}_{3}^{i} \nu^{c}_{i}$, again eq.(\ref{eq:P_1}). If we don't label the Froggatt-Nielsen messengers in the internal lines, and omit other field insertions, we have figure \ref{fig:P_1}.
Starting with eq.(\ref{eq:P_@}), eq.(\ref{eq:M_123}) is the only choice, and we conclude with eq.(\ref{eq:P_@}) again, in figure \ref{fig:P_@}.
Starting with eq.(\ref{eq:P_o}), eq.(\ref{eq:M_23}) is the only choice, and we conclude with eq.(\ref{eq:P_o}) again, in figure \ref{fig:P_o}.

Each diagram repeats the same $\nu_{i}$ contraction in both external lines, so with figure \ref{fig:P_1} decoupling (due to eq.(\ref{eq:M_hier}), despite eq.(\ref{eq:Y_hier})),
we conclude that
$\nu_{@} \equiv \langle \bar{\phi}_{23}^{i} \rangle \nu_{i}$ (figure \ref{fig:P_@}) and
$\nu_{\odot} \equiv \langle \bar{\phi}_{123}^{i} \rangle \nu_{i}$ (figure \ref{fig:P_o}) are mass eigenstates.
As $\lambda_{23} \left| b \right| ^{2} > \lambda_{123} \left| c \right| ^{2}$,
the heaviest state is $\nu_{@}$. In the flavon VEV basis, $\nu_{@} \propto \nu_{2}-\nu_{3}$ has equal parts of the second and third families, and the lighter $\nu_{\odot} \propto \nu_{1}+\nu_{2}+\nu_{3}$ is equal parts of all three families.
We have exact neutrino TBM mixing (the lightest state is the orthogonal combination).

While we won't specify the charged lepton masses, we note that in SUSY GUT models with underlying $SO(10)$ (like \cite{Ivo1,Ivo3}) the charged leptons have small mixing (the charged lepton mass terms are shared by the down quarks, see \cite{Ivo0} for more on the corrections to TBM mixing).
If the model is embedded in this type of SUSY GUT, charged lepton mixing won't deviate leptonic mixing far from TBM.
$\nu_{@}$ is then the atmospheric neutrino state and $\nu_{\odot}$ the solar state, hence the subscript labels (the framework leads to a normal hierarchy, but we avoid the $\nu_{3}$ and $\nu_{2}$ notation of these states due to our use of the family index subscript). The resulting leptonic mixing will be close to eq.(\ref{eq:TBM}) and thus within the current experimental bounds.

\subsection{Effective symmetry}

Another way to understand how the $P_{Y}$ and $P_{M}$ terms combine to give neutrino TBM mixing is to notice that the terms actually preserve an effective symmetry \footnote{This reasoning, suggested by Graham Ross, was already used in \cite{Ivo0, Ross:2007zz}.} that, after type I seesaw, leads to the effective neutrino TBM Lagrangian $L_{\nu}$:

\begin{equation}
L_{\nu} = \Lambda_{@} \nu_{@}^{2} +  \Lambda_{\odot} \nu_{\odot}^{2}
\label{eq:effL}
\end{equation}
The $\Lambda$ are related with the respective $\lambda$ and to the appropriate Majorana masses.

$L_{\nu}$ gives TBM mixing as there is no mixing between $\nu_{@}$ and $\nu_{\odot}$. Consider a $Z_{2}$ symmetry under which
$\bar{\phi}_{23} \rightarrow - \bar{\phi}_{23}$, which keeps the unwanted $\bar{\phi}_{23}^{i} \nu_{i} \bar{\phi}_{123}^{j} \nu_{j}$ term out of eq.(\ref{eq:effL}). Under this $Z_{2}$ the $P_{Y}$ terms in eq.(\ref{eq:P_@}) and eq.(\ref{eq:P_o}) are invariant, if we have also (for example) $\nu \rightarrow - \nu$ (still preserving $L_{\nu}$).
Eq.(\ref{eq:P_1}) is not invariant under these assignments, but that won't matter as long as the largest Majorana mass decouples $\bar{\phi}_{3}^{i} \nu_{i}$ from $L_{\nu}$.
With these charge assignments, $P_{Y}$ can't have terms with a $\bar{\phi}_{23}$ pair or with a $\bar{\phi}_{123}$ pair.
In contrast, the Majorana terms are invariant if they include either two or zero $\bar{\phi}_{23}$, as in eq.(\ref{eq:M_23}) and eq.(\ref{eq:M_123}).
These are precisely the framework's key properties that $P_{Y}$ and $P_{M}$ must possess in order to produce TBM for the effective neutrinos  (in complete models the effective $Z_{2}$ symmetry results from the symmetries keeping those unwanted terms out of $P_{Y}$ and $P_{M}$).

\subsection{Explicit diagonalisation}

The neutrino mass terms in $P_{Y}$ give rise to the Dirac neutrino mass matrix $M_{D}$:

\begin{equation}
M_{D} =
\left( 
\begin{array}{ccc}
0 & \lambda_{\odot} bc & -\lambda_{\odot} bc \\ 
\lambda_{@} bc & (\lambda_{@} + \lambda_{\odot}) bc & (\lambda_{@} -\lambda_{\odot}) bc \\ 
-\lambda_{@} bc & (-\lambda_{@} + \lambda_{\odot}) bc & aa \lambda_{1} + (-\lambda_{@} -\lambda_{\odot}) bc
\end{array}
\right)
\label{eq:D_mat}
\end{equation}
As $\lambda_{@} = \lambda_{\odot}$, $M_{D}$ is symmetric.

The mass terms in $P_{M}$ give rise to the Majorana neutrino mass matrix $M_{M}$:

\begin{equation}
M_{M} =
\left( 
\begin{array}{ccc}
\lambda_{123} cc & \lambda_{123} cc & \lambda_{123} cc \\ 
\lambda_{123} cc & \lambda_{23} bb + \lambda_{123} cc & -\lambda_{23} bb+ \lambda_{123} cc \\ 
\lambda_{123} cc & -\lambda_{23} bb + \lambda_{123} cc & \lambda_{3} aa + \lambda_{23} bb + \lambda_{123} cc
\end{array}
\right)
\label{eq:M_mat}
\end{equation}
The matrices are presented in the flavon VEV basis, where they are easy to write as the flavon VEVs are the (easy to remember) eq.(\ref{eq:P3 vev}), eq.(\ref{eq:P23 vev}) and eq.(\ref{eq:P123 vev}). We can input $M_{D}$ and $M_{M}$ into the type I seesaw formula:

\begin{equation}
m_{\nu} = - M_{D} M_{M}^{-1} M_{D}^{T}
\label{eq:Iseesaw}
\end{equation}
$m_{\nu}$ can then be diagonalised, and the eigenvectors must correspond to neutrino TBM mixing when there is sufficient hierarchy in the Majorana masses. It is perhaps more pedagogical to take the large $\lambda_{3} \left| a \right| ^{2}$ limit before diagonalisation, enabling an explicit analytical calculation of the eigenvectors. In this limit, the third column of $M_{D}$ decouples from the effective neutrinos, so we can consider the simplified form $M_{D}'$ (despite $M_{D}^{33}$ being larger than the other entries):

\begin{equation}
M_{D}' =
\left( 
\begin{array}{ccc}
0 & D & 0 \\ 
D & 2 D & 0 \\ 
-D & 0 & 0
\end{array}
\right)
\label{eq:D'_mat}
\end{equation}
We defined $D \equiv \lambda_{@} bc \equiv \lambda_{\odot} bc$. In the same limit, we can use $M_{M}'$, a block diagonal form of $M_{M}$, obtained by neglecting the small off-diagonal entries of the third row and column (compared to the very large $M_{M}^{33}$):

\begin{equation}
M_{M}' = 
\left( 
\begin{array}{ccc}
\lambda_{123} cc & \lambda_{123} cc & 0 \\ 
\lambda_{123} cc & \lambda_{23} bb + \lambda_{123} cc & 0 \\ 
0 & 0 & M_{3}
\end{array}
\right)
\label{eq:M'_mat}
\end{equation}
$M_{3}$ is the (very large) mass of the heaviest right-handed neutrino, originating from the term in eq.(\ref{eq:M_3}).

Within the approximation, we compute the effective neutrino mass matrix $m_{\nu}'$:

\begin{equation}
m_{\nu}' = - M_{D}' M_{M}'^{-1} M_{D}'^{T}=
- \frac{D^{2}}{\lambda_{123} cc}
\left( 
\begin{array}{ccc}
0 & 0 & 0 \\ 
0 & 1 & -1 \\ 
0 & -1 & 1
\end{array}
\right)
-\frac{D^{2}}{\lambda_{23} bb}
\left( 
\begin{array}{ccc}
1 & 1 & 1 \\ 
1 & 1 & 1 \\ 
1 & 1 & 1
\end{array}
\right)
\label{eq:mn'_mat}
\end{equation}
Diagonalising $m_{\nu}'$ is trivial, and results in $m_{@}=2 \left| D^{2}/ (\lambda_{123} cc) \right| > m_{\odot}=3 \left| D^{2}/ (\lambda_{23} bb) \right|$ as the two non-zero eigenvalues, corresponding respectively to the TBM eigenstates $\nu_{@}$ and $\nu_{\odot}$.

\section{Summary}

We highlighted the key properties of the framework of neutrino mass terms used in \cite{Ivo1, Ivo3}, illustrating it with simple terms in $P_{Y}$ and $P_{M}$.
We then explained how those terms interact through the type I seesaw mechanism and, if there is enough hierarchy in the Majorana masses to result in sequential dominance, give rise to effective neutrino masses that correspond to exact neutrino TBM.

\section*{Acknowledgments}
We are grateful to Graham Ross and Stefano Morisi for helpful discussions.

\noindent All diagrams were created with JaxoDraw \cite{Jaxo}.

\noindent The work of I. de Medeiros Varzielas was supported by FCT under the grant
SFRH/BPD/35919/2007.

\bibliography{refs}
\bibliographystyle{hepv}

\end{document}